\documentclass{article}
\usepackage{spconf,amsmath,amssymb,graphicx,booktabs,array,tabularx,hyperref,multirow}
\hypersetup{hidelinks}
\newcommand{\Lcal}{\mathcal{L}}
\newcommand{\E}{\mathbb{E}}
\newcommand{\sg}{\operatorname{sg}}

\title{ROUTE-DMD: DISTRIBUTION MATCHING DISTILLATION WITH\\TASK-ROUTED MECHANISM FOR SPEECH FOUNDATION MODELS}
\name{\shortstack{\itshape Yanru Huo$^{1,2}$, Zhikang Niu$^{1,3}$, Wenming Tu$^{1,2}$, Junxi Liu$^{1,3}$, Tian Tan$^{1}$,\\
\itshape Shan Yang$^{5}$, Yipeng Kang$^{2}$, Ziyang Ma$^{1,3,4}$, Xie Chen$^{1,3,*}$, Zilong Zheng$^{2,*}$}}

\address{$^{1}$ MoE Key Lab of Artificial Intelligence, X-LANCE Lab, Shanghai Jiao Tong University \\ $^{2}$ State Key Laboratory of General Artificial Intelligence, BIGAI \\
$^{3}$ Shanghai Innovation Institute, $^{4}$ Nanyang Technological University, 
$^{5}$ Tencent Hunyuan}

\begin{document}
\ninept
\maketitle

\begingroup\renewcommand{\thefootnote}{*}\footnotetext[0]{Corresponding author}\endgroup

\begin{abstract}
Speech foundation models built on flow matching handle generation, editing, and processing in one system, but every request costs 20--50 iterative steps, and classifier-free guidance (CFG) doubles each of them. Distribution matching distillation (DMD) trains a student that approaches the teacher's quality in a few steps, yet existing speech distillation work applies it to generation alone. We show that a distribution-matching objective does not serve every task: in separation, the noised states of erroneous student outputs deviate far from the teacher's training distribution, where its prediction is unreliable. Route-DMD therefore makes the objective task-dependent, routing the affected tasks to a regression loss and keeping decoupled distribution matching elsewhere. The student is initialized by consistency distillation, and teacher rollouts use adaptive projected guidance (APG) instead of CFG. Without CFG, the 4-step student matches the 32-step teacher on generation, editing, and processing tasks at a 4.5$\times$ end-to-end speedup\footnote{Demo: \url{https://route-dmd.github.io/}. Code: \url{https://github.com/iHateTheWorld555/AuK-RouteDMD}}.

\end{abstract}

\begin{keywords}
speech foundation model, distribution matching distillation, few-step sampling, speech synthesis, speech editing
\end{keywords}

\section{Introduction}
\label{sec:intro}

\begin{figure*}[!t]
\centering
\includegraphics[width=0.85\textwidth]{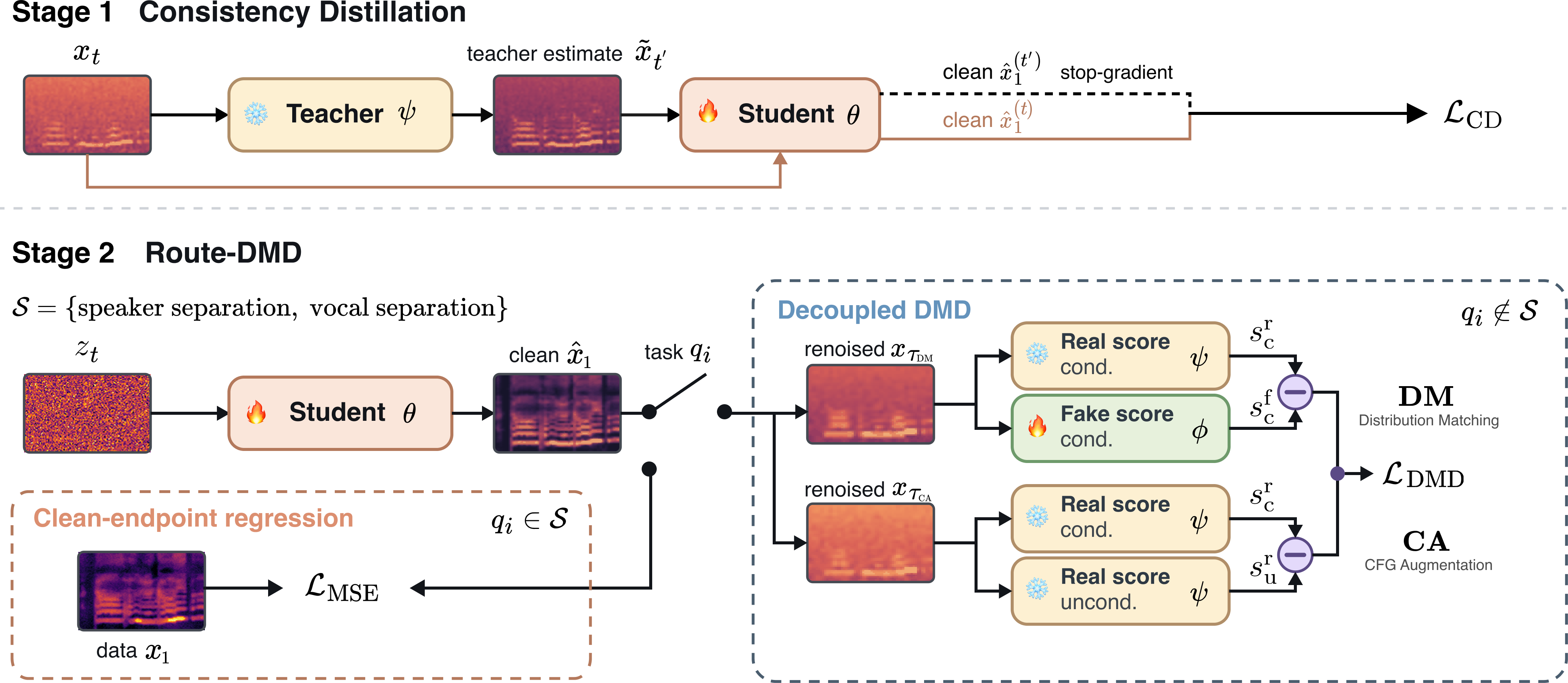}
\caption{Overview of Route-DMD. \textbf{Stage 1} uses consistency distillation for initialization. \textbf{Stage 2} routes separation samples to a regression loss and other samples to Decoupled DMD.}
\label{fig:overview}
\end{figure*}

Diffusion and flow-matching models have achieved high-quality generation across image~\cite{ZImage}, video~\cite{mova,wan2025wan}, and audio domains~\cite{F5TTS,megatts}. In speech, this approach now supports not only zero-shot TTS but also speech foundation models that combine generation and editing in one system, with AuK~\cite{auk} pushing this direction by integrating up to 15 speech tasks into one model with competitive performance. However, flow-matching models face an efficiency bottleneck: each request typically requires 20--50 neural function evaluations, and CFG doubles the computation per step. This significantly increases response latency during deployment.

Few-step distillation addresses this bottleneck by replacing the teacher with a student model that samples in a few steps. These methods can be categorized into trajectory-level approaches~\cite{CM,MeanFlow}, which are stable but limited by the teacher's trajectory, and distribution-level approaches~\cite{DMD2,DecoupledDMD}, which achieve higher few-step fidelity. Speech systems have adopted both approaches~\cite{SplitMeanFlow,DMOSpeech2}; however, all of them are validated solely on generation tasks. The distillation of a speech foundation model that also supports editing and speech processing remains unexplored.

Applying DMD to a speech foundation model, however, severely degrades its separation capability. In this task, the noised state of an erroneous student sample deviates far from the distribution where the teacher is trained, making the teacher's predictions unreliable. Allowing the teacher to predict from a noised ground-truth target mitigates this degradation (Fig.~\ref{fig:routing-ablation}). Additionally, since separation has a single correct answer given the instruction, a per-sample loss can reliably supervise the student against an exact target.

To address this gap, we propose \emph{Route-DMD} (Fig.~\ref{fig:overview}), a two-stage distillation framework for speech foundation models. The first stage uses consistency distillation to give the student model with a few-step starting point for the subsequent DMD stage, leading to faster convergence and improved final generation quality. In the second stage, training is routed by task: separation samples are trained with a regression loss and excluded from DMD, while generation and editing samples, including TTS and instruction-guided speech editing, are trained with Decoupled DMD~\cite{DecoupledDMD}. Additionally, we replace CFG with APG~\cite{APG} in the DMD stage to prevent the artifacts of CFG-guided teacher predictions from propagating to the student.

Our contributions are threefold. First, we show that distribution matching does not serve every task, and attribute the failure to unreliable teacher predictions. Second, we propose Route-DMD, which routes separation tasks to regression loss and initializes the student with consistency distillation. Third, we adopt APG teacher rollouts to prevent guidance artifacts during distillation. The student model requires only four steps without CFG and matches or surpasses the 32-step teacher with a 4.5$\times$ end-to-end speedup.

\section{Related Work}
\label{sec:related}

\subsection{From Task-Specific Models to Speech Foundation Models}

Task-specific models have been developed for speech synthesis~\cite{F5TTS,megatts}, editing~\cite{Voicecraft,megatts2}, enhancement~\cite{re-use}, separation~\cite{MossFormer2}, and super-resolution~\cite{audiosr}. Recent unified models combine generation and editing~\cite{MingUniAudio,StepAudioEditX}, with some also incorporating speech processing~\cite{QuarkAudio}. AuK~\cite{auk} unifies generation, editing, and processing within a single framework, which takes a further step toward unified speech foundation models.

\subsection{Acceleration of Flow-Matching Models}

Acceleration methods for flow-matching models can be categorized into two approaches. \textbf{1) Trajectory-level} methods train a student to reproduce the teacher's ODE trajectory~\cite{CM,MeanFlow}, with speech-specific variants following~\cite{SplitMeanFlow, liang2025towards}. The training is stable, but the student's fidelity is bounded by the teacher's trajectory. \textbf{2) Distribution-level} methods instead match the teacher's output distribution~\cite{DMD,DMD2,DMOSpeech2} and achieve higher few-step fidelity, while Decoupled DMD~\cite{DecoupledDMD} decomposes the distillation gradient further (Sec.~\ref{sec:preliminaries}). Neither approach has been applied to a speech foundation model. 

\section{Route-DMD}
\label{sec:method}

\subsection{Preliminaries}\label{sec:preliminaries}
\textbf{Flow matching.}
We follow the optimal-transport conditional flow-matching formulation~\cite{FlowMatching}. Clean target latents follow the distribution $p_1(x_1)$, the prior is $p_0(x_0)=\mathcal{N}(x_0;0,I)$, and $c$ denotes the conditioning inputs, including the text instruction and the input audio. For $t\sim\mathcal{U}(0,1)$, the interpolant $x_t=(1-t)x_0+t x_1$ induces a conditional probability path, whose target vector field is the constant $x_1-x_0$. We train the parameterized vector field $v_\theta$ with
\begin{equation}
 \Lcal_{\mathrm{CFM}}(\theta)=\E_{t,x_0,x_1,c}
 \!\left[\left\|v_\theta\!\left(x_t,t,c\right)-\left(x_1-x_0\right)\right\|^2\right].
 \label{eq:flow_path}
\end{equation}
At inference, the ODE $\mathrm{d}x_t/\mathrm{d}t=v_\theta(x_t,t,c)$ is integrated over $t\in[0,1]$ from $x_0\sim p_0$. With CFG, the sampling velocity is $\widetilde v=v_{\mathrm{u}}+\alpha(v_{\mathrm{c}}-v_{\mathrm{u}})$, where $v_{\mathrm{c}}$ and $v_{\mathrm{u}}$ are the conditional and unconditional velocities and $\alpha$ is the guidance scale.

\noindent \textbf{Decoupled DMD.} The student $f_\theta$ is initialized from the teacher and inherits the velocity parameterization of Eq.~\eqref{eq:flow_path}. To build its input, the student rolls out from $z_0\sim p_0$ and stops at a randomly drawn time $t$, yielding $z_t$. The student then predicts $\hat x_1=f_\theta(z_t,t,c)$. From the same student output, we independently sample $\tau_{\mathrm{DM}}\sim\mathcal{U}(0,1)$ for the distribution matching (DM) term and $\tau_{\mathrm{CA}}\sim\mathcal{U}(t,1)$ for the CFG augmentation (CA) term, then add Gaussian noise to $\hat{x}_1$ to obtain $x_{\tau_{\mathrm{DM}}}$ and $x_{\tau_{\mathrm{CA}}}$, following the schedules of~\cite{DecoupledDMD}. The frozen multi-step teacher serves as the \emph{real score} model, and an online auxiliary network $f_\phi$, trained on the student's outputs, serves as the \emph{fake score} model. Both predict velocity, and we write $s$ for the score they induce, a reweighted form of the velocity. For a latent $x$, the real score model gives the conditional and unconditional predictions $s_{\mathrm{c}}^{\mathrm{r}}(x)$ and $s_{\mathrm{u}}^{\mathrm{r}}(x)$, and the fake score model gives the conditional prediction $s_{\mathrm{c}}^{\mathrm{f}}(x)$. With $\alpha$ denoting the teacher's guidance scale, the DM and CA score differences define the student gradient as
\begin{multline}
 \nabla_\theta\Lcal_{\mathrm{DMD}}=\E\Big[-\Big(
 s_{\mathrm{c}}^{\mathrm{r}}(x_{\tau_{\mathrm{DM}}})-s_{\mathrm{c}}^{\mathrm{f}}(x_{\tau_{\mathrm{DM}}})\\[-2pt]
 +(\alpha-1)\big(s_{\mathrm{c}}^{\mathrm{r}}(x_{\tau_{\mathrm{CA}}})-s_{\mathrm{u}}^{\mathrm{r}}(x_{\tau_{\mathrm{CA}}})\big)\Big)\partial_\theta\hat x_1\Big].
 \label{eq:dmd}
\end{multline}

\begin{table*}[!ht]\centering\setlength{\tabcolsep}{2.6pt}\renewcommand{\arraystretch}{1.0}\setlength{\aboverulesep}{0.25ex}\setlength{\belowrulesep}{0.25ex}\caption{Comparison with prior systems and the teacher on representative tasks. The best results are in \textbf{bold}, and the second-best results are \underline{underlined}.}\label{tab:main}\begin{tabularx}{\textwidth}{@{}>{\centering\arraybackslash}p{0.18\textwidth}p{0.20\textwidth}>{\centering\arraybackslash}p{0.18\textwidth}>{\centering\arraybackslash}X@{}}\toprule\textbf{Dataset} & \textbf{System} & \textbf{Metrics} & \textbf{Results} \\\midrule\multirow{4}{*}{\shortstack{\textbf{SeedTTS-Eval}\\ en\textbar{}zh\textbar{}zh-hard\textbar{}avg}} &Qwen3-TTS~\cite{qwen3-tts} &\multirow{4}{*}{SIM$\uparrow$ / WER$\downarrow$} &0.717 / 1.23 \textbar{} 0.770 / 1.22 \textbar{} 0.748 / 6.76 \textbar{} 0.745 / 3.07 \\& VoxCPM 2~\cite{VoxCPM2} && 0.753 / 1.84 \textbar{} 0.795 / \textbf{0.97} \textbar{} 0.753 / 8.13 \textbar{} 0.767 / 3.65 \\& AuK (32 steps) && \textbf{0.788} / \textbf{1.02} \textbar{} \textbf{0.814} / \underline{1.02} \textbar{} \underline{0.782} / \textbf{5.91} \textbar{} \textbf{0.795} / \textbf{2.65} \\& \quad w/ Route-DMD (4 steps) && \underline{0.772} / \underline{1.03} \textbar{} \textbf{0.814} / 1.10 \textbar{} \textbf{0.784} / \underline{6.43} \textbar{} \underline{0.790} / \underline{2.85} \\\midrule\multirow{3}{*}{\shortstack{\textbf{InstructTTSEval}\\ DSD-ZH\textbar{}DSD-EN}} &Qwen3-TTS-VD &\multirow{3}{*}{ACC$\uparrow$} &\underline{81.1} \textbar{} \textbf{82.4} \\& AuK (32 steps) && \textbf{83.4} \textbar{} \underline{81.6} \\& \quad w/ Route-DMD (4 steps) && 78.8 \textbar{} \textbf{82.4} \\\midrule\multirow{4}{*}{\textbf{MMAE-Speech}} &Step-Audio-EditX~\cite{StepAudioEditX} &\multirow{4}{*}{IFR$\uparrow$ / CR$\uparrow$ / EMR$\uparrow$} &43.52 / 77.27 / 4.69 \\& Ming-UniAudio~\cite{MingUniAudio} && 34.13 / 76.01 / 7.04 \\& AuK (32 steps) && \textbf{48.23} / \textbf{88.11} / \underline{12.44} \\& \quad w/ Route-DMD (4 steps) && \underline{46.62} / \underline{86.41} / \textbf{13.85} \\\midrule\multirow{3}{*}{\textbf{CHiME-4}} &RE-USE~\cite{re-use} &\multirow{3}{*}{OVRL$\uparrow$ / WER$\downarrow$ / UTMOS$\uparrow$} &\underline{3.33} / 10.71 / 3.44 \\& AuK (32 steps) && 3.28 / \underline{7.98} / \underline{3.72} \\& \quad w/ Route-DMD (4 steps) && \textbf{3.35} / \textbf{7.84} / \textbf{3.91} \\\midrule\multirow{3}{*}{\textbf{Libri2Mix}} &MossFormer2-SS~\cite{MossFormer2} &\multirow{3}{*}{OVRL$\uparrow$ / WER$\downarrow$ / UTMOS$\uparrow$} &3.24 / \underline{9.34} / 3.66 \\& AuK (32 steps) && \underline{3.28} / \textbf{9.12} / \underline{3.87} \\& \quad w/ Route-DMD (4 steps) && \textbf{3.32} / 10.07 / \textbf{4.03} \\\midrule\multirow{3}{*}{\textbf{VCTK-SR}} &AudioSR~\cite{audiosr} &\multirow{3}{*}{OVRL$\uparrow$ / WER$\downarrow$ / UTMOS$\uparrow$} &3.08 / 4.72 / 3.13 \\& AuK (32 steps) && \underline{3.21} / \underline{3.06} / \underline{3.93} \\& \quad w/ Route-DMD (4 steps) && \textbf{3.25} / \textbf{2.92} / \textbf{4.05} \\\bottomrule\end{tabularx}\end{table*}

\subsection{Consistency Initialization}

Initializing the student with consistency distillation allows the DMD stage to converge in fewer updates and reach higher final quality, consistent with observations in video generation~\cite{CausVid,DistillAlign}. In the flow-matching convention of Sec.~\ref{sec:preliminaries}, for adjacent times $t<t'$, the frozen teacher advances $x_t$ to the cleaner time $t'$ with a one-step CFG velocity:

\begin{equation}
 \begin{aligned}
  \tilde{x}_{t'}&=x_t+(t'-t)\widetilde v(x_t,t,c),\\[-2pt]
  \widetilde v&=v_{\mathrm{u}}+\alpha(v_{\mathrm{c}}-v_{\mathrm{u}}),
 \end{aligned}
\end{equation}
The student maps the noisy samples at both times to clean endpoints, and we train it with the following two-point consistency objective:
\begin{equation}
  \Lcal_{\mathrm{CD}}(\theta)=\E\!\left[\frac{\bigl\|f_\theta(x_t,t,c)-\sg\!\left(f_\theta(\tilde{x}_{t'},t',c)\right)\bigr\|_2}{t'-t}\right],
 \label{eq:cm}
\end{equation}
where $\sg(\cdot)$ denotes the stop-gradient operator.

\subsection{Task-routed Mechanism}
Eq.~\eqref{eq:dmd} supervises the student through the real score model, which Eq.~\eqref{eq:flow_path} trains only on the path from noise to ground-truth targets. This supervision breaks down during separation. When an erroneous separation output is noised, the resulting state deviates significantly from that training distribution, making the real score model's prediction there is unreliable. Replacing the student output with a ground-truth target before noising, relieves this degradation (Fig.~\ref{fig:routing-ablation}). Furthermore, unlike generation, separation has a single correct answer once the instruction is given, so a regression loss can supervise the student against that answer without the real score model. Let $q_i$ be the task label of sample $i$ and $\mathcal{S}$ the set of routed tasks, comprising speaker and vocal separation. Samples in $\mathcal{S}$ receive a regression loss and are excluded from DMD training, while all other samples are trained with Decoupled DMD:
\begin{equation}
 (\Lcal_{\theta}^{(i)},\Lcal_\phi^{(i)})=
 \begin{cases}
  (\|\hat x_{1,i}-x_{1,i}\|^{2},\,0), & \text{if }q_i\in\mathcal{S},\\[-1pt]
  (\Lcal_{\mathrm{DMD}}^{(i)},\Lcal_{\mathrm{fake}}^{(i)}), & \text{otherwise}.
 \end{cases}
 \label{eq:routing}
\end{equation}
Here, $\hat x_{1,i}$ is the student's clean-endpoint prediction, $x_{1,i}$ is the ground-truth clean latent, $\Lcal_{\mathrm{DMD}}^{(i)}$ is the per-sample Decoupled DMD loss, and $\Lcal_{\mathrm{fake}}^{(i)}$ is the online flow-matching loss for the fake score network $f_\phi$. Task-specific supervision preserves separation capability without sacrificing generation quality.

\subsection{Adaptive Projected Guidance}
CFG at high guidance scales over-extrapolates along the conditional direction, causing artifacts \cite{APG}. Moreover, during DMD training, the student absorbs this bias from CFG-guided teacher rollouts. We therefore use APG~\cite{APG} in place of CFG in the teacher rollouts of the DMD stage. APG operates in clean-prediction space and decomposes the conditional-minus-unconditional residual with respect to the conditional prediction, keeping the orthogonal component and down-weighting the parallel component by $\eta$. Reducing this parallel component suppresses the artifacts while retaining guidance.

\section{Experiments}
\label{sec:experiments}

\subsection{Experimental Setup}

\textbf{Training details.}
We validate our method on AuK. The student and the fake score model are trained with AdamW, a constant learning rate of $10^{-5}$, gradient clipping at 1.0, and a total batch size of 2.56 million audio frames on the same data used to train AuK. Consistency initialization starts the student from the teacher and trains it for 500 updates with teacher CFG at guidance scale 2. Route-DMD then starts both the student and the fake score model from that checkpoint, with a $5:1$ ratio of fake-score to student updates and no adversarial loss. The teacher estimates the real score with APG at guidance scale 4 and $\eta=0$.

\noindent \textbf{Evaluation details.} Our evaluations span generation, editing, and processing. For generation, SeedTTS-Eval~\cite{SeedTTS} scores zero-shot TTS by word error rate (WER), measured with Whisper-large-v3~\cite{Whisper} for English and Paraformer-zh~\cite{funasr} for Chinese, and speaker similarity (SIM), computed as the cosine similarity of WavLM-large-based speaker embeddings~\cite{wavlm}. The Chinese and English descriptive-style directive (DSD) subsets of InstructTTSEval~\cite{InstructTTSEval} score instruction TTS by accuracy (ACC). For editing, the speech subset of MMAE~\cite{MMAE} gives the instruction following rate (IFR), consistency rate (CR), and exact match rate (EMR). For processing, CHiME-4~\cite{CHiME4} covers enhancement, Libri2Mix~\cite{LibriMix} separation, and VCTK-SR~\cite{VCTK} super-resolution, each under DNSMOS P.835 overall quality (OVRL)~\cite{DNSMOS}, UTMOS~\cite{UTMOS}, and WER. The end-to-end real-time factor (RTF) is measured on the test-en subset of SeedTTS-Eval times inference. Subjective evaluation reports five-point mean opinion scores (MOS) for naturalness (N-MOS) and instruction adherence (I-MOS).

\begin{figure}[!t]
\centering
\includegraphics[width=0.85\linewidth]{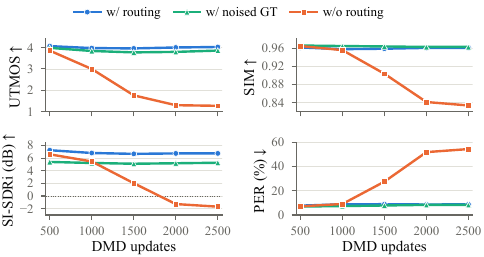}
\caption{Ablation of the task-routed mechanism on Libri2Mix. Separation metrics vs.\ Decoupled DMD
updates for the regression loss (w/ routing), Decoupled DMD with the teacher queried on a noised
ground-truth target rather than a noised student output (w/ noised GT), and plain Decoupled DMD
(w/o routing).}
\label{fig:routing-ablation}
\end{figure}

\begin{table}[!t]
\centering
\setlength{\tabcolsep}{6pt}
\renewcommand{\arraystretch}{1.0}
\caption{Results of the subjective evaluation are reported as mean ± 95\% confidence interval (CI) over all ratings. End-to-end RTF is measured on a single NVIDIA H20 GPU.}
\label{tab:subjective}
\begin{tabular}{@{}lccc@{}}
\toprule
System & N-MOS$\uparrow$ & I-MOS$\uparrow$ & RTF$\downarrow$ \\
\midrule
AuK (32 steps) & 4.23 \(\pm\) 0.10 & 4.50 \(\pm\) 0.09 & 0.4333 \\
\quad w/ Route-DMD (4 steps) & 4.34 \(\pm\) 0.10 & 4.57 \(\pm\) 0.09 & \textbf{0.0958} \\
\bottomrule
\end{tabular}
\end{table}

\subsection{Main Results}
\textbf{The student retains broad capability while improving efficiency.}
With only four steps, the student remains competitive with AuK on generation and editing, while achieving the best SIM of 0.784 on the test-zh-hard subset and the best EMR of 13.85 in the speech subset of MMAE (Table~\ref{tab:main}). In speech processing, it exceeds AuK on OVRL and UTMOS across all three benchmarks and achieves the best results on every metric of CHiME-4 and VCTK-SR. The student matches AuK in both N-MOS and I-MOS within the reported 95\% CIs while reducing RTF from 0.4333 to 0.0958, a 4.5$\times$ speedup (Table~\ref{tab:subjective}). These results show that the accelerated student preserves its task coverage.

\subsection{Ablation Studies}
\label{sec:ablation}
\begin{figure}[!t]
\centering
\includegraphics[width=0.85\linewidth]{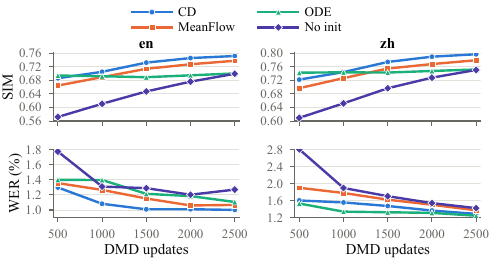}
\caption{Ablation of initialization on SeedTTS-Eval. Each curve corresponds to a student initialized before DMD by consistency distillation (CD), ODE regression (ODE), meanflow distillation, or no initialization.}
\label{fig:ablation}
\end{figure}
\begin{figure}[!t]
\centering
\includegraphics[width=0.85\linewidth]{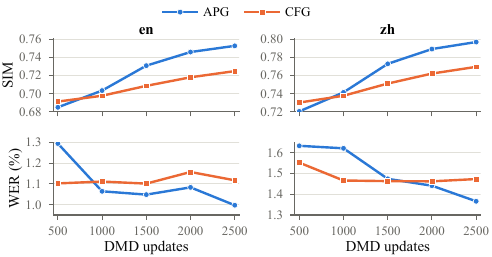}
\caption{Ablation of guidance in the DMD teacher rollouts. SIM and WER on SeedTTS-Eval versus DMD updates.}
\label{fig:apg}
\end{figure}

\noindent \textbf{Consistency initialization provides the strongest few-step starting point.}
Fig.~\ref{fig:ablation} compares four ways of initializing the student before DMD, matched in update budget and teacher guidance strength: consistency distillation (CD), meanflow distillation, ODE regression, and no initialization. The three initialized variants differ only in the objective of this stage. Consistency initialization leads on nearly every SIM and WER curve and converges to the best final quality. ODE regression initialization reaches a competitive Chinese WER but shows little change in SIM. Without initialization, the model lags behind in both WER and SIM from early training onward, and the gap does not close. Consistency initialization therefore provides the most stable starting point.

\noindent \textbf{The task-routed mechanism preserves separation capability under DMD.} To expose the collapse at the signal level, we report the scale-invariant signal-to-distortion ratio improvement (SI-SDRi), which measures how much the output improves over the input mixture, and the phone error rate (PER), which measures the separated stream's purity. Fig.~\ref{fig:routing-ablation} shows that the routed student remains stable, with UTMOS, SIM, SI-SDRi, and PER staying near 4.0, 0.96, +6.7 dB, and 8.8\%, respectively, throughout training. Without routing, SI-SDRi drops from +6.6 to -1.7 dB and PER rises from 7.1\% to 54.4\%, alongside sharp declines in UTMOS and speaker similarity. Feeding the teacher a noised ground-truth target in place of the noised student output keeps DMD on these samples yet avoids the collapse, holding SI-SDRi at +5.1 to +5.4 dB and PER within 7.0 to 8.4\%, which locates the failure in the input the teacher receives. However, routing still ends higher on SI-SDRi and UTMOS, which makes it the better choice.

\noindent \textbf{APG improves the final quality ceiling of DMD training.}
We compare CFG and APG teacher rollouts in the DMD stage at the same guidance scale $\alpha=4$. With CFG, the student improves faster initially but soon saturates, whereas with APG the student continues to improve and finishes with higher SIM and lower WER in both languages (Fig.~\ref{fig:apg}). APG therefore yields the better student.

\section{Conclusion}
\label{sec:conclusion}
This work identifies a task-dependent limitation of DMD when distilling a unified speech foundation model. The analysis suggests that, for separation, noised student outputs can deviate from the real-score model's training distribution, leading to unreliable supervision. Route-DMD addresses this by routing separation to direct regression on paired ground-truth targets while retaining Decoupled DMD for the remaining tasks. The framework also incorporates consistency initialization and APG-guided teacher rollouts in place of CFG. Experimental results show that the resulting 4-step student achieves performance comparable to the 32-step teacher on generation and editing, with higher DNSMOS OVRL and UTMOS scores across all three speech-processing benchmarks and a 4.5$\times$ end-to-end speedup. These findings highlight the value of task-aware supervision for preserving the speech foundation model's broad capabilities during few-step distillation.

\section{Compliance with Ethical Standards}
The subjective listening test is conducted with the informed consent of all participants, who are compensated for their time. This study is performed in line with the principles of the Declaration of Helsinki.

\bibliographystyle{IEEEbib}
\bibliography{strings,refs}

\end{document}